 \documentclass[final,3p,times,twocolumn]{elsarticle}

\usepackage{graphicx}

\usepackage{amssymb}
\usepackage{amsmath}

\usepackage{xtab}

\usepackage{booktabs,caption,fixltx2e}
\usepackage[flushleft]{threeparttable}
\usepackage{color}

\usepackage[nodots,nocompress]{numcompress}

\newcommand{\mbbf}[1]{\mbox{\boldmath${#1}$}}

\newcommand{\cra}[1]{\textcolor{black}{#1}}
\newcommand{\crb}[1]{\textcolor{black}{#1}}
\newcommand{\crc}[1]{\textcolor{black}{#1}}

\biboptions{square}

\journal{Acta Astronautica}

\begin{document}

\begin{frontmatter}


\ead{petri.toivanen@fmi.fi}
\fntext[]{{\it Telephone number}: +358-50-5471521}

\title{Thrust vectoring of an electric solar wind sail with a
  realistic sail shape}


\author{P. Toivanen and P. Janhunen}

\address{Finnish Meteorological Institute, FIN-00101, Helsinki, Finland}

\begin{abstract}

The shape of a rotating electric solar wind sail under the centrifugal
force and solar wind dynamic pressure is modeled to address the sail
attitude maintenance and thrust vectoring. The sail rig assumes
centrifugally \crb{stretched} main tethers that extend radially outward
from the spacecraft in the sail spin plane. Furthermore, the tips of
the main tethers host remote units that are connected by auxiliary
tethers at the sail rim. Here, we derive the equation of main tether
shape and present both a numerical solution and an analytical
approximation for the shape as parametrized both by \crb{the ratio of
  the electric sail force to the centrifugal force} and the sail
orientation with respect to the solar wind direction. The resulting
shape is such that near the spacecraft, the roots of the main tethers
form a cone, whereas towards the rim, this coning is flattened by the
centrifugal force, and the sail is coplanar with the sail spin
plane. Our approximation for the sail shape is parametrized only by
the tether root coning angle and the main tether length. Using the
approximate shape, we obtain the torque and thrust of the electric
sail force applied to the sail. As a result, the amplitude of the
tether voltage modulation required for the maintenance of the sail
attitude is given as a torque-free solution. The amplitude is smaller
than that previously obtained for a rigid single tether resembling a
spherical pendulum. This implies that less thrusting margin is
required for the maintenance of the sail attitude. For a given voltage
modulation, the thrust vectoring is then considered in terms of the
radial and transverse thrust components.

\end{abstract}

\begin{keyword}

Electric solar wind sail \sep Attitude control \sep Transverse thrust


\end{keyword}

\end{frontmatter}

\section*{Nomenclature}
\noindent\begin{xtabular}{@{}lcl@{}}
a &=& voltage modulation torque-free\\
c &=& cosine function\\
${\bf e}$ &=& unit vector\\
${\bf F}$ &=& electric sail force\\
${\boldsymbol{\mathcal{F}}}$ &=& total sail thrust\\
${\bf G}$ &=& centrifugal force\\
$g$ &=& voltage modulation general\\
$\mathcal{I}$ &=& integral\\
$k$ &=& force ratio\\
$L$ &=& main tether length\\
$l$ &=& coordinate along the main tether\\ 
$M$ &=& total mass\\
$N$ &=& number of main tethers\\
$m$ &=& single main tether mass\\
s &=& sine function\\
${\bf T}$ &=& main tether tension\\
${\boldsymbol{\mathfrak{T}}}$ &=& electric sail torque\\
${\boldsymbol{\mathcal{T}}}$ &=& total sail torque\\
$u$ &=& local tether tangent\\
${\bf v}$ &=& solar wind velocity\\
$v$ &=& solar wind speed\\
$(x,y,z)$ &=& Cartesian coordinates\\
$\alpha$ &=& sail angle\\
$\gamma$ &=& local tether coning angle\\
$\Delta t$ &=& rotation period\\
$\mu$ &=& linear mass density\\
$\psi$ &=& thrust angle\\
$(\rho,\phi,z)$ &=& circular cylindrical coordinates\\
${\mbbf \tau}$ &=& angular torque density\\
$\xi$ &=& electric sail force factor\\
$\omega$ &=& sail spin rate\\
\end{xtabular} \\

\textit{Subscripts} \\
\noindent\begin{xtabular}{@{}lcl@{}}
$0$ &=& tether root\\
$i$ &=& index\\
$L$ &=& tether length\\
${\rm mt}$ &=& main tether\\
$q$ &=& vector component index\\
${\rm ru}$ &=& remote unit\\
${\rm s}$ &=& sail\\
$(x,y,z)$ &=& Cartesian coordinates\\
$\alpha$ &=& sail angle\\
$\gamma$ &=& local tether coning angle\\
$(\rho,\phi,z)$ &=& circular cylindrical coordinates\\
\end{xtabular}\\

\textit{Superscripts} \\
\noindent\begin{xtabular}{@{}lcl@{}}
$j$ &=& summation index\\
$*$ &=& \crb{orbital frame of reference}
\end{xtabular}


\section{Introduction}
\label{sec:intro}

\begin{figure}[htb]
	\centering\includegraphics[width=2.6in]{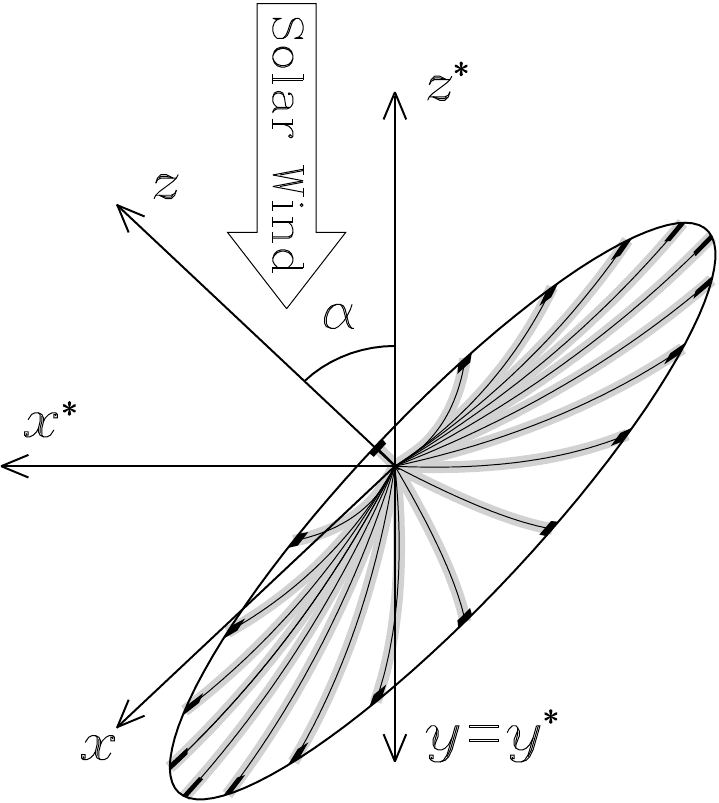}
	\caption{Electric sail flight configuration and \crb{two coordinate
          systems: $(x^{\ast},y^{\ast},z^{\ast})$ is the orbital frame
          of reference; and $(x,y,z)$ is rotated around the $y^{\ast}$
          axis by the sail angle alpha.}}
	\label{fig:sail_schema}
\end{figure}

The electric solar wind sail is a propulsion system that uses the
solar wind proton flow as a source of momentum for spacecraft thrust
\cite{eka}. The momentum of the solar wind is transferred to the
spacecraft by electrically charged light-weight tethers that deflect
the proton flow. The sail electrostatic effective area is then much
larger than the mechanical area of the tethers, and the system
promises high specific acceleration \crc{up to about 10 mm/s$^2$}
\cite{rsi}. As the tethers are polarized at a high positive voltage
they attract electrons that in turn tend to neutralize the tether
charge state. However, only a modest amount of electric power \crc{of
  a few hundred watts} is required to operate electron guns to
maintain the sail charge state, and the sail can easily be powered
by solar panels \cite{janhunensandroos,toka}. The main tethers are
centrifugally deployed radially outward from the spacecraft in the
sail spin plane (Fig. \ref{fig:sail_schema}). To be tolerant to the
micro-meteoroid flux each tether has a redundant structure that
comprises a number (typically 4) of 20-50 $\mu$m metal wires bonded to
each other, for example by ultrasonic welding \cite{onekmtether}. As a
baseline design, the tips of the main tethers host remote units that
are connected by auxiliary tethers at the sail perimeter to provide
mechanical stability to the sail\cite{fp7}.

As the electric sail offers a large effective sail area with modest
power consumption and low mass, it promises a propellantless
continuous low thrust system for spacecraft propulsion for various
kinds of missions \cite{applications}. These include fast transit to
the heliopause \cite{outersolarsystem}, missions in non-Keplerian
orbit such as helioseismology in a solar halo orbit
\cite{nonkeplerian}, space weather monitoring with an extended warning
time (closer to the sun than L1), multi-asteroid touring
mission. \crb{Using the electric sail, such missions can typically be
  accomplished without planetary gravity assist maneuvers and
  associated launch windows. If planetary swing-bys are planned during
  the mission, each solar eclipse has to be carefully considered to
  avoid drastic thermal contraction and expansion of the sail tethers
  \cite{eclipse}.} In addition to scientific missions, the electric
sail can be used for planetary defense as a gravity tractor
\cite{gravitytractor} or an impactor \cite{impactor} and to rendezvous
with such Potentially Hazardous Objects that cannot be reached by
conventional propulsion systems \cite{pho}. The electric sail has also
been suggested as a key \crb{method of transportation for products of
  asteroid mining}\crc{\cite{ky26}}. \crb{Specifically, water from
  asteroids can be used} for in-orbit production of LH2/LOX by
electrolysis to provide a cost efficient way of transporting
infrastructure associated with manned Mars missions \cite{emmi}.

\cra{The electric sail has an intrinsic means for its flight control,
  i.e., spin plane attitude control, maintenance, and maneuvers. These
  can be realized by applying differential voltage modulation to the
  sail tethers synchronously with the sail spin \cite{controla}. Thus
  the flight control is similar to the helicopter rotor flight control
  based on the blades' angle of attack.} Furthermore, the sail can
fully be turned off for orbital coasting phases or proximity maneuvers
near light weight targets such as small asteroids. The coasting phases
are also central to optimal transfer orbits between circular, for
example, planetary orbits \cite{marsorbit} (when reaching a target in
an elliptical orbit such as the comet 67P/Churyumov-Gerasimenko
coasting phases are not needed \crc{\cite{rosetta}}). \crb{Note that these
  coasting phases are not associated with the planetary gravity assist
  maneuvers}. Navigation to the target is also feasible, in spite of
the variable nature of the solar wind \cite{navigation}.

In this paper, we derive an integral equation for the sail main tether
shape under the solar wind dynamical pressure and the centrifugal
forces in Sec. \ref{subsec:eqshape}. The resulting equation of the
tether shape is then solved numerically (Sec. \ref{subsec:numerical})
and an analytical approximation for the shape is then obtained
(Sec. {\ref{subsec:approximate}). Using this approximation, we obtain
  general expressions for the thrust (Sec. \ref{subsec:thrust}) and
  the torque (Sec. \ref{subsec:torque}) arising from the solar wind
  transfer of momentum to the sail. In Sec. \ref{subsec:torquefree},
  we introduce a tether voltage modulation that leads to a torque-free
  sail motion. Finally, in Sec. \ref{subsec:thrustvectoring}, we
  consider the sail thrust vectoring in terms of both the radial and
  transverse thrust.

\cra{The reference frames used in this paper are illustrated in
  Fig. \ref{fig:sail_schema}. One of the frames
  $(x^{\ast},y^{\ast},z^{\ast})$ is the orbital reference frame with
  the $z^{\ast}$ axis pointing to the sun, the $y^{\ast}$ axis being
  in the direction of the negative normal of the orbital plane, and
  the $x^{\ast}$ completing the triad in the direction of the orbital
  velocity vector. In the other system $(x,y,z)$, $z$ is aligned with
  the sail spin axis, and $x$ is chosen so that the solar wind nominal
  direction is in the $xz$ plane. These two systems are related by a
  rotation around $y^{\ast}$ axis by the sail angle $\alpha$. In the
  $xyz$ system, the circular cylindrical coordinates ($\rho,\phi,z$)
  are used.} 

\crb{The reference frames introduced above are local in the following
  sense: they rotate with respect to the distant stars while the sail
  is orbiting around the sun; however, the sail itself keeps its
  orientation with respect to the distant stars; and thus the sail
  spin axis is slowly rotating (360$^{\circ}$/yr) in these
  non-inertial local frames in terms of the Coriolis effect. In order
  to maintain the sail orientation with respect to the sun, an
  additional tether voltage modulation has to be introduced. The
  amplitude of this modulation is, however, much smaller compared to
  the modulation associated with the inclined sail\cite{controla}, and
  the Coriolis effect can be neglected in this work. It is noted,
  however, that the Coriolis effect can only be partially canceled by
  the main tether voltage modulation and it leads to a secular
  variation in the sail spin rate\cite{controla}. This is a topic
  considered in a future study that addresses the electric sail spin
  rate variations and control using the model developed in this
  paper.}

\section{Tether shape}
\label{sec:tethershape}

\subsection{Equation of tether shape}
\label{subsec:eqshape}

The electric sail tether shape under the solar wind forcing can be
obtained by writing an integral equation similar to that of a
catenary\cite{catenary}. Fig. \ref{fig:shape_schema} shows the
electric sail force and the centrifugal force influencing the tether
shape. Local unit vectors parallel and perpendicular to the tether can
be written \crb{in terms of sine and cosine of the local coning angle
  $\gamma$} as
\begin{eqnarray}
{\bf e}_{\parallel} &=& {\rm c}_{\gamma}\,{\bf e}_{\rho} + {\rm s}_{\gamma}{\bf e}_z\label{eq:eparallel}\\
{\bf e}_{\perp} &=& {\rm s}_{\gamma}{\bf e}_{\rho} - {\rm c}_{\gamma}{\bf e}_z.
\label{eq:eperpendicular}
\end{eqnarray}
\crb{According to Fig. \ref{fig:shape_schema}, the total force ${\bf
    T} = {\bf F} + {\bf G}$ that equals the tether tension can be
  split into} $\rho$ and $z$ components as
\begin{equation}
\frac{T_z}{T_{\rho}} = \tan\gamma = \frac{dz}{d\rho} \equiv u(\rho),
\end{equation}
\crb{where we have introduced the local tether tangent $u(\rho)$}. An
equation for the tether shape can then simply be written as
\begin{equation}
u = \frac{F_z}{G+F_{\rho}}.
\label{eq:impeqofshape}
\end{equation}
Note that the forces present here are the total forces integrated over
the tether from the reference point $\rho$ to the tether tip at
$\rho_L$. 
\begin{figure}[htb]
	\centering\includegraphics[width=3.0in]{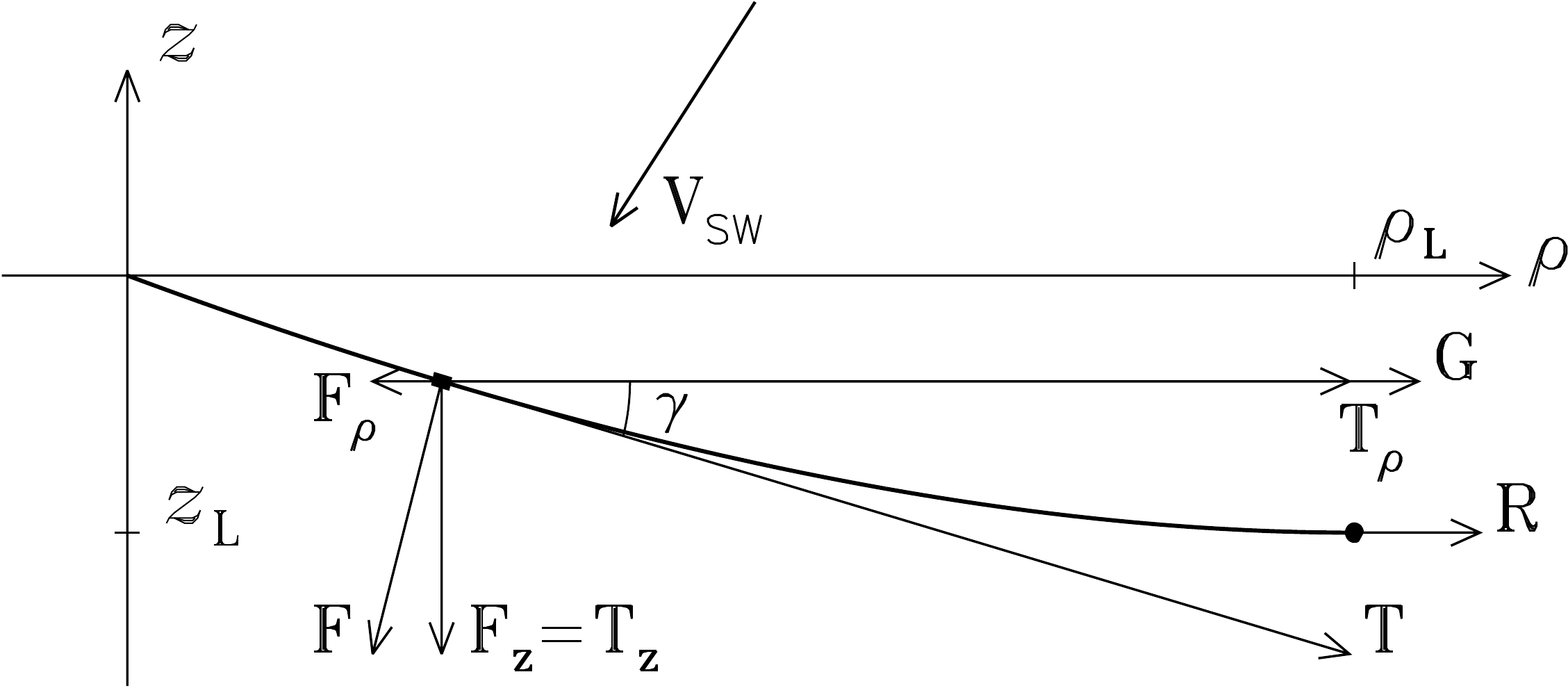}
	\caption{Electric sail tether (thick solid curve), remote unit
          (black dot).}
	\label{fig:shape_schema}
\end{figure}

For a tether segment $dl$ with a mass of $dm_{\rm mt}$, the centrifugal force
($dG = \omega^2\rho dm_{\rm mt}$) can be written in terms of the tether linear
mass density $\mu$ ($dG = \mu\omega^2\rho dl$). As the length of the
tether segment reads as
\begin{equation}
dl = \sqrt{1+\left(\frac{dz}{d\rho}\right)^2}d\rho = \sqrt{1+u^2}\;d\rho,
\label{eq:linesegment}
\end{equation}
the total centrifugal force is
\begin{equation}
G = \mu\omega^2\int_{\rho}^{\rho_L}\rho\sqrt{1+u^2}\;d\rho + m_{\rm ru}\omega^2\rho_L,
\label{eq:totcentri}
\end{equation}
where the last term is the centrifugal force exerted by the remote
unit including the auxiliary tether mass.

The electric sail force per unit tether length is directed along the
solar wind velocity component perpendicular to the tether direction as
\begin{equation}
\frac{d{\bf F}}{dl} = \xi{\bf v}_{\perp}
\label{eq:forcelaw}
\end{equation}
where ${\bf v}_{\perp}$ is the solar wind component perpendicular to
the main tether direction and $\xi$ is a force factor arising from the
electric sail thrust law \cite{janhunensandroos}. Similarly to the
centrifugal force above, the electric sail force can be integrated to
give
\begin{equation}
{\bf F} = \int_{\rho}^{\rho_L}\xi{\bf v}_{\perp}\sqrt{1+u^2}d\rho.
\label{eq:esailforce}
\end{equation}
\crc{As the solar wind velocity is assumed to be radial, it can be written as}
\begin{equation}
{\bf v} = v(s_{\alpha}{\bf e}_{\rho} + c_{\alpha}{\bf e}_z)
\label{eq:solarwind}
\end{equation}
\crc{in terms of the sail angle and solar wind speed with typical
  values of about 400 km/s}. The component perpendicular to the tether
direction can be expressed in terms of the unit vector of
Eq. (\ref{eq:eperpendicular}) as
\begin{eqnarray}
{\bf v}_{\perp} &=& ({\bf v}\cdot{\bf e}_{\perp}){\bf e}_{\perp}\nonumber\\
&=& v({\rm s}_{\alpha}{\rm s}_{\gamma}^2-{\rm c}_{\alpha}{\rm s}_{\gamma}{\rm c}_{\gamma}){\bf e}_{\rho} + v({\rm c}_{\alpha}{\rm c}_{\gamma}^2-{\rm s}_{\alpha}{\rm s}_{\gamma}{\rm c}_{\gamma}){\bf e}_z.
\label{eq:vperp}
\end{eqnarray}
Using trigonometric identities to express ${\rm s}_{\gamma}$ and
${\rm c}_{\gamma}$ in terms of $\tan\gamma$
(with $\tan\gamma = u$), $\rho$ and $z$ components of the electric sail
force (\ref{eq:esailforce}) can be written as
\begin{equation}
F_{\rho} = - \xi v\int_{\rho}^{\rho_L}\frac{({\rm c}_{\alpha} - {\rm s}_{\alpha} u)u}{\sqrt{1+u^2}}\;d\rho
\label{eq:totesailrho}
\end{equation}
and
\begin{equation}
F_z = \xi v\int_{\rho}^{\rho_L}\frac{{\rm c}_{\alpha} - {\rm s}_{\alpha} u}{\sqrt{1+u^2}}\;d\rho
\label{eq:totesailz}
\end{equation}
Finally, inserting the integral force terms in
Eq. (\ref{eq:impeqofshape}), the equation of shape of the tether can be
written as
\begin{equation}
u = \frac{\xi v\int_{\rho}^{\rho_L}\frac{{\rm c}_{\alpha} - {\rm s}_{\alpha} u}{\sqrt{1+u^2}}\;d\rho}{\mu\omega^2\int_{\rho}^{\rho_L}\rho\sqrt{1+u^2}\;d\rho + m_{\rm ru}\omega^2\rho_L- \xi v\int_{\rho}^{\rho_L}\frac{({\rm c}_{\alpha} - {\rm s}_{\alpha} u)u}{\sqrt{1+u^2}}\;d\rho}
\label{eq:expeqofshape}
\end{equation}
In addition, the tether extent in $\rho$, $\rho_L$ is determined by
the tether length and shape as
\begin{equation}
L = \int_{\rho_0}^{\rho_L}{\sqrt{1+u^2}}\;d\rho.
\label{eq:lengthintegral}
\end{equation}
The shape of the tether can then be solved using
Eqs. (\ref{eq:expeqofshape}) and (\ref{eq:lengthintegral}).

\subsection{Numerical solution}
\label{subsec:numerical}

Numerical solution to Eq. (\ref{eq:expeqofshape}) can be found by
considering $z(\rho)$ being locally linear as $z_i = u_i \rho + c_i$
at $\rho = \rho_i$. All integrals in Eq. (\ref{eq:expeqofshape})
depend only on $u$ and $\rho$, and we are left to find a recurrence
relation only for $u_i$. To do so, an integral $\mathcal{I}$ of any
general function $h(\rho,u)$ can be written as
\begin{equation}
\mathcal{I}_i = \int_{\rho_i}^{\rho_L} h(\rho,u)d\rho = h(\rho_i,u_i)\Delta\rho_i + \mathcal{I}_{i-1}.
\label{eq:integral}
\end{equation}
An equation for $u_i$ can be obtained by substituting all integrals in
Eq. (\ref{eq:expeqofshape}) with Eq. (\ref{eq:integral}), accordingly. After some algebra, $u_i$ can be written as
\begin{equation}
u_i = \frac{\xi v {\rm c}_{\alpha} \Delta L + F_{i-1}^z}{(\xi v {\rm s}_{\alpha} + \mu_t\omega^2\rho_{i-1})\Delta L + G_{i-1} + m_R\omega^2 \rho_L - F_{i-1}^{\rho}}.
\label{eq:ui}
\end{equation}
Given an initial starting point $\rho_L$, a numerical solution can be
found recursively using Eq. (\ref{eq:ui}) over the tether length. As
$\rho_L$ is unknown, depending on the initial guess of $\rho_L$, the
process is iterated until the solved tether root distance equals the
actual tether attachment point at the
spacecraft. Fig. \ref{fig:solution_demo} shows the tether shape
$z(\rho)$ and the local tether tangent $u$. \cra{Parameter values used are
$L$ = 20 km, $\xi v$ = 0.5 mN/km, $\alpha$ = $45^{\circ}$, $\mu$ = 10
g/km, $m_{\rm ru}$ = 1 kg, $\Delta t$ = 125 min. These values are
motivated as follows: a baseline sail assumes hundred tethers with a
length of 20 km each; the thrust per tether length of 0.5 mN/km
translates to a baseline thrust of 1 N; tether linear mass density is
about 10 g/km \cite{onekmtether}; a remote unit with a dry mass of
about 0.5 kg was developed and qualification tested in an EU/FP7/ESAIL
project \cite{fp7}; and the rotation period of 125 min is used here
for a prominent tether coning to visualize the tether shape.} Note that
the solution can be easily verified by calculating the force
integrals in Eq. (\ref{eq:expeqofshape}) as shown in bottom panel of
Fig. \ref{fig:solution_demo} and equating these against $u$ as in
Eq. (\ref{eq:expeqofshape}).
\begin{figure}[htb]
	\centering\includegraphics[width=2.7in]{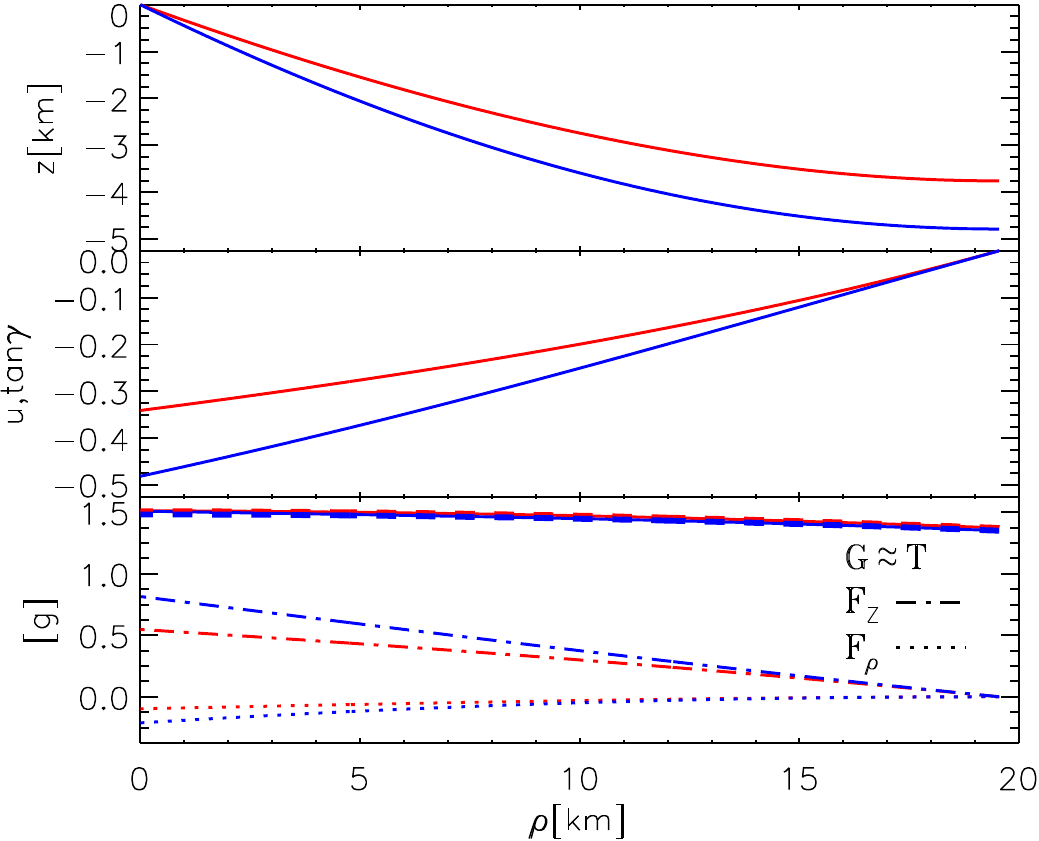}
	\caption{Tether shape (top), tether tangent (middle), and the
          force terms of the equation of tether shape (bottom) for a
          slowly rotating sail with low tether tension of 1.5
          grams. Red (blue) curve corresponds to the tether azimuth
          angle, $\phi = 0$ ($\phi = \pi$).}
	\label{fig:solution_demo}
\end{figure}

\subsection{Analytical approximation}
\label{subsec:approximate}

An analytical approximation for the tether shape can be obtained for
a weakly coning sail ($u \approx 0$). \cra{Fig. \ref{fig:solution_real} shows
the numerically obtained tether shape with a maximum tether tension of
5 grams. As the tether \crb{can tolerate a tension of about 13 grams}
at maximum \cite{onekmtether}, the tension of 5 grams leaves a clear safety
margin to 13 grams.} The parameter values are the same as in
Fig. \ref{fig:solution_demo} except the sail spin is faster, and the
rotation period, $\Delta t$ = 70 min. In general, an approximation for
the equation of shape (\ref{eq:expeqofshape}) can be found as an
expansion of $\rho = b_0 + b_1 u + b_2 u^2$. After solving the
coefficients ($b_0,b_1,b_2$) using Eqs. (\ref{eq:expeqofshape}) and
(\ref{eq:lengthintegral}), $u$ can be solved from the expansion
above. However, for the purposes of this paper we simplify the
analysis and consider only the linear terms so that $u$ can be written
as
\begin{equation}
u = u_0\left(1-\frac{\rho}{\rho_L}\right)
\label{eq:linearu}
\end{equation}
As it can be seen in Fig. \ref{fig:solution_real}, this is well
justified, and $u = u_0$ at $\rho = 0$ and $u = 0$ at $\rho = \rho_L$
as it is the case. The tether shape can then be integrated ($dz/d\rho
= u$) to give
\begin{equation}
z = u_0\rho\left(1-\frac{\rho}{2\rho_L}\right).
\label{eq:linearz}
\end{equation}
\begin{figure}[htb]
	\centering\includegraphics[width=2.7in]{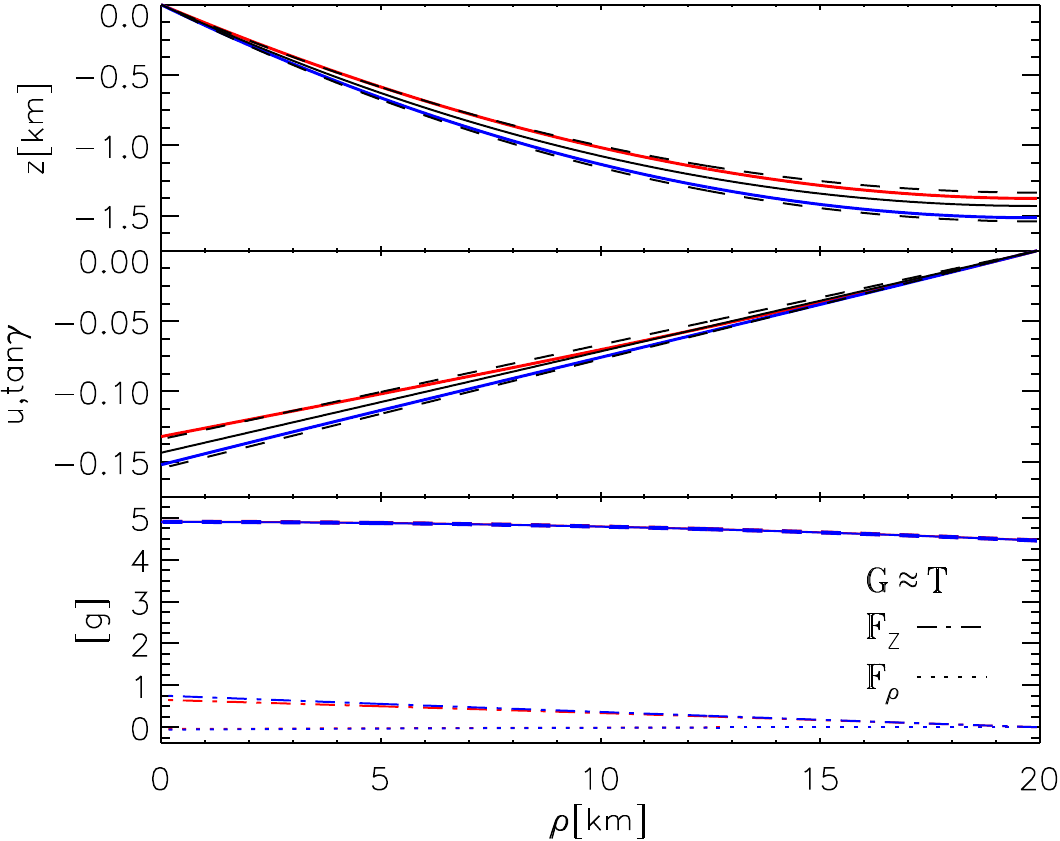}
	\caption{Tether shape (top), tether tangent (middle), and the
          force terms of the equation of tether shape (bottom) for a
          sail with maximum tether tension of 5 grams. Red (blue) curve
          corresponds to the tether azimuth angle, $\phi = 0$ ($\phi =
          \pi$). Dashed lines shows the corresponding analytical
          approximations and black line is the sail shape.}
	\label{fig:solution_real}
\end{figure}

To finalize our model for the tether shape we are left to solve
$\rho_L$ and $u_0$ as functions of the sail and solar wind
parameters. Using Eq. (\ref{eq:lengthintegral}), expanding
$\sqrt{1+u^2}$ as a power series in $u$, and integrating, $\rho_L$ can
be expressed in terms of the total tether length as
\begin{equation}
\rho_L = L\left(1-\frac{1}{6}u_0^2\right)
\label{eq:rhoL}
\end{equation}
The equation of shape (\ref{eq:expeqofshape}) at $\rho = 0$ can be written as
\begin{equation}
u_0 = \frac{\xi v\int_{0}^{\rho_L}({\rm c}_{\alpha} - {\rm s}_{\alpha} u)\;d\rho}{\mu\omega^2\int_{0}^{\rho_L}\rho\;d\rho + m_r\omega^2\rho_L - \xi v\int_{0}^{\rho_L}{\rm c}_{\alpha}u\;d\rho}
\label{eq:appeqofshape}
\end{equation}
by excluding terms higher than first order in $u_0$ ($\sqrt{1+u^2}
\approx 1$). Noting that $\int_{0}^{\rho_L}u\;d\rho = \rho_L
u_0/2$, one can solve $u_0$ to obtain
\begin{equation}
u_0 = \frac{2k\cos\alpha}{2 + k\sin\alpha},
\label{eq:u0}
\end{equation}
where
\begin{equation}
k = \frac{2 \xi v}{(m_{\rm mt} + 2 m_{\rm ru})\omega^2}
\label{eq:k}
\end{equation}
is the ratio of the electric sail force to the centrifugal
force. Fig. \ref{fig:solution_real} shows the approximations for the
shape for the sail angles of -$\alpha$ and $\alpha$ corresponding to
the tether azimuth locations of $\phi = 0$ and $\phi = \pi$,
respectively.

\subsection{Sail shape}

The shape of the model sail \crb{is parametrized by} the radial extent
of the sail ($\rho_{\rm s}$) and the tangent of the sail coning angle
($u_{\rm s}$) at the spacecraft. The sail radial extent is
trivial and it equals the single tether length up to second order in
$u_{\rm s}$ as in Eq. \ref{eq:rhoL}, and we are left to only determine
$u_{\rm s}$.

Here, we present two estimates for $u_{\rm s}$ based on the results shown
above. One solution is to use Eq. (\ref{eq:u0}) to give the sail
coning tangent as an average of tether tangents at $\pm \alpha$,
\begin{equation}
u_{\rm s} = \frac{4k\cos\alpha}{4-k^2\sin^2\alpha}.
\label{eq:usa}
\end{equation}
The other solution is to consider the solar wind vector to be rotated
around the $z$ axis in sail coordinates to the locations of the
individual tethers. Then, as the solar wind components in the sail
plane cancel when averaging over the tethers, we are left with an
effective solar wind $z$ component $v_{\rm eff} = v\cos\alpha$. Then,
using Eq. (\ref{eq:u0}) with the zero effective sail angle, the sail
coning tangent is given as
\begin{equation}
u_{\rm s} = k_{\rm eff} = k\cos\alpha.
\label{eq:usb}
\end{equation}
As the centrifugal force is typically much larger than the electric
sail force ($k \ll 1$), Eqs. (\ref{eq:usa}) and (\ref{eq:usb}) are
essentially equal.

\section{Sail thrust and torque}
\label{sec:rigid}

\subsection{Thrust}
\label{subsec:thrust}

The total sail thrust is calculated by summing over the
number of tethers ($N$) and integrating over the single tethers as
\begin{equation}
\mathcal{F}_q = \sum_{j=1}^N\int_0^L\frac{dF_q^j}{dl}dl
\label{eq:deftotaltorque}
\end{equation}
By changing variables ($l \rightarrow \rho \rightarrow u$), the integral in Eq. (\ref{eq:deftotaltorque}) can be written as
\begin{equation}
\mathcal{F}_q = \sum_{j=1}^N\int_0^{u_{\rm s}}\frac{\rho_L}{u_{\rm s}}\frac{dF_q^j}{dl}\sqrt{1+u^2}du
\end{equation}
Next, we assume that the sail comprises such a large number of tethers
(, i.e., $N \gtrsim 12$) that the summation over the tethers in
Eq. (\ref{eq:deftotaltorque}) can be replaced by integration over the
tether azimuthal locations in $\phi$ as
\begin{equation}
\sum_{j=1}^NF(\phi_j) \rightarrow N\int_0^{2\pi}f(\phi)d\phi, 
\label{eq:sumtoint}
\end{equation}
where $f(\phi) = F(\phi)/{2\pi}$ can be considered as the angular thrust
density. The total thrust is then an integral of the thrust
density and it can be written as
\begin{equation}
\mathcal{F}_q = N\int_0^{2\pi}\int_0^{u_{\rm s}}\frac{\rho_L}{u_{\rm s}}\frac{df_q}{dl}\sqrt{1+u^2}dud\phi.
\end{equation}
According to the electric sail force law of Eq. (\ref{eq:forcelaw}),
the thrust on a line segment $dl$ is given as
\begin{equation}
\frac{d{\bf F}}{dl} = g_{\phi}\xi{\bf v}_{\perp},
\label{eq:dforcedl}
\end{equation}
where we have added the tether voltage modulation $g_{\phi}$. The
modulation is scaled to the maximum voltage with $g_{\phi} \in
[0,1]$. We also assume for simplicity that the solar wind velocity is
given as
\begin{equation}
{\bf v} = v_x{\bf e}_x + v_z{\bf e}_z.
\end{equation}
Its component perpendicular to the tether reads then as
\begin{eqnarray}
{\bf v}_{\perp} &=& {\bf v} - ({\bf v}\cdot{\bf e}_{\parallel}){\bf e}_{\parallel}\nonumber\\
&=& {\bf v} - (v_x {\rm c}_{\gamma}{\rm c}_{\phi } + v_z {\rm s}_{\gamma}){\bf e}_{\parallel},
\label{eq:swvperp}
\end{eqnarray}
where the unit vector parallel to the tether is given by ${\bf
  e}_{\parallel} = {\rm c}_{\gamma}{\bf e}_{\rho} + {\rm
  s}_{\gamma}{\bf e}_z$ as in Eq. (\ref{eq:eparallel}). Since ${\bf
  e}_{\rho} = {\rm c}_{\phi}{\bf e}_x + {\rm s}_{\phi}{\bf e}_y$ in
the circular cylindrical coordinate system, the thrust components per
line segment can be expressed as
\begin{eqnarray}
\frac{d{F_x}}{dl} &=& g_{\phi}\xi[v_x - (v_x {\rm c}_{\gamma}{\rm c}_{\phi } + v_z {\rm s}_{\gamma}){\rm c}_{\gamma}{\rm c}_{\phi}] \nonumber \\
\frac{d{F_y}}{dl} &=& -g_{\phi}\xi(v_x {\rm c}_{\gamma}{\rm c}_{\phi } + v_z {\rm s}_{\gamma}){\rm c}_{\gamma}{\rm s}_{\phi}\nonumber \\
\frac{d{F_z}}{dl}  &=& g_{\phi}\xi[v_z - (v_x {\rm c}_{\gamma}{\rm c}_{\phi } + v_z {\rm s}_{\gamma}){\rm s}_{\gamma}].
\label{eq:dforcedl}
\end{eqnarray}
The next step is to integrate over the tether length, i.e., from zero
to $u_{\rm s}$ in terms of $u$. \crc{Using the shape of the sail
  tethers as given by Eq. (\ref{eq:linearu}) with $u_0 = u_{\rm s}$ we
  determine the thrust to the second order in $u_{\rm s}$}. This can
be accomplished by using any computer algebra system such as Maxima
\cite{maxima}, and the angular thrust density can be given as
\begin{eqnarray}
f_x &=& \frac{g_{\phi}\xi L}{2\pi} \left[v_x - \frac{1}{2}v_zu_{\rm s}{\rm c}_{\phi } - v_x\left(1-\frac{1}{3}u_{\rm s}^2\right){\rm c}_{\phi }^2\right] \nonumber \\
f_y &=& -\frac{g_{\phi}\xi L}{2\pi} \left[\frac{1}{2}v_z u_{\rm s}{\rm s}_{\phi } + v_x\left(1-\frac{1}{3}u_{\rm s}^2\right){\rm s}_{\phi }{\rm c}_{\phi }\right]\nonumber \\
f_z  &=& \frac{g_{\phi}\xi L}{2\pi} \left[v_z\left(1-\frac{1}{3}u_{\rm s}^2\right)-\frac{1}{2}v_xu_{\rm s}{\rm c}_{\phi }\right].
\label{eq:totalforce}
\end{eqnarray}
Note that to obtain the total force to the entire sail
Eq. (\ref{eq:totalforce}) has to be integrated over the sail in $\phi$
for a given voltage modulation. In Sec. \ref{subsec:thrustvectoring},
this will be done for the modulation that results in torque-free sail
dynamics.

\subsection{Torque}
\label{subsec:torque}

By definition, the torque on a tether segment $dl$ generated by the
electric sail force Eq. (\ref{eq:dforcedl}) is given as
\begin{equation}
\frac{d\mathfrak{T}_q}{dl} = g_{\phi}\xi\left[{\bf r}\times{\bf v}_{\perp}\right]_q.
\label{eq:esailtorque}
\end{equation}
Writing ${\bf v}_{\perp}$ as in Eq. (\ref{eq:swvperp}) and ${\bf r} =
\rho {\rm c}_{\phi}{\bf e}_x + \rho {\rm s}_{\phi}{\bf e}_y + z{\bf
  e}_z$, the cross product ${\bf r}\times{\bf v}_{\perp}$ can be
calculated and the torque per line segment can be written as
\begin{eqnarray}
\frac{d\mathfrak{T}_x}{dl} &=& g_{\phi}\xi[\rho v_z {\rm s}_{\phi} - (v_x {\rm c}_{\gamma}{\rm c}_{\phi } + v_z {\rm s}_{\gamma})(\rho {\rm s}_{\gamma}-z {\rm c}_{\gamma}){\rm s}_{\phi}] \nonumber \\
\frac{d\mathfrak{T}_y}{dl} &=& g_{\phi}\xi[z v_x - \rho v_z {\rm c}_{\phi} + (v_x {\rm c}_{\gamma}{\rm c}_{\phi } + v_z {\rm s}_{\gamma})(\rho {\rm s}_{\gamma}-z {\rm c}_{\gamma}){\rm c}_{\phi}]\nonumber \\
\frac{d\mathfrak{T}_z}{dl}  &=& -g_{\phi}\xi\rho v_x {\rm s}_{\phi}.
\label{eq:dtorquedl}
\end{eqnarray}
The angular torque density can then be obtained by integration over
the tether length as in Eq. (\ref{eq:totalforce}), and the torque
density reads as
\begin{eqnarray}
\tau_x &=& \frac{g_{\phi}\xi L^2}{4\pi}\left[v_z\left(1-\frac{1}{6}u_{\rm s}^2\right){\rm s}_{\phi} + \frac{1}{3} v_x u_{\rm s}{\rm c}_{\phi}{\rm s}_{\phi}\right]\nonumber \\
\tau_y &=& \frac{g_{\phi}\xi L^2}{4\pi}\left[\frac{2}{3}v_x u_{\rm s} - v_z\left(1-\frac{1}{6}u_{\rm s}^2\right){\rm c}_{\phi} - \frac{1}{3}v_x u_{\rm s}{\rm c}_{\phi}^2\right]\nonumber \\
\tau_z  &=& -\frac{g_{\phi}\xi L^2}{4\pi} v_x\left(1-\frac{1}{4}u_{\rm s}^2\right){\rm s}_{\phi}.
\label{eq:totaltorque}
\end{eqnarray}
Note that Eq. (\ref{eq:totaltorque}) has to be integrated over the sail in
$\phi$ for a given voltage modulation to obtain the total sail torque.

\section{Results}
\label{sec:results}

\subsection{Torque-free sail dynamics}
\label{subsec:torquefree}

In order to find torque-free dynamics for the sail, we apply
a modulation given as
\begin{equation}
g_{\phi} = 1-a(1 \pm {\rm c}_{\phi}).
\label{eq:modulation}
\end{equation}
where $\pm$ corresponds to $\pm \alpha$. After integrating
Eq. (\ref{eq:totaltorque}), \cra{only the $y$ component of the total
  torque is different from zero and it can be expressed as}
\begin{equation}
\mathcal{T}_y = \frac{1}{4}N\xi L^2\left[v_x u_{\rm s} - a\left(v_x u_{\rm s} \mp (v_z - \frac{1}{6} v_z u_{\rm s}^2)\right)\right].
\end{equation}
Setting $\mathcal{T}_y$ equal to zero, the amplitude $a$ can be solved and it is
seen that with the modulation given in
Eq. (\ref{eq:modulation}), the sail dynamics is free of torque when
\begin{equation}
a = - u_{\rm s}\tan\alpha\left(1 + u_{\rm s}\tan\alpha + \mathcal{O}(u_{\rm s}^2)\right), 
\label{eq:modulation_amplitude}
\end{equation}
where $v_x/v_z$ is replaced with $\pm\tan\alpha$. For a non-inclined
($\alpha = 0^{\circ}$) or fully planar ($u_{\rm s} = 0$) sail, the
efficiency equals 1 as no voltage modulation is needed for the sail
attitude control. Otherwise, a portion of the available voltage is
required for the sail control which decreases the sail efficiency as
shown in Fig. \ref{fig:efficiency}. \cra{Here, the efficiency of the
  tether voltage modulation, and below, the rest of the results are
  shown as contour plots as a function of the sail angle and the ratio
  of the electric sail force to the centrifugal force as given in
  Eq. (\ref{eq:k})}. Note that the second order terms in
Eq. (\ref{eq:modulation_amplitude}) and expressions below are given in
Tab. \ref{tab:2ndorder} merely as estimates for the validity of the
power series expansions, and any geometric interpretations based on
these terms are conceivably irrelevant.

\begin{table}
\centering
\begin{threeparttable}
\caption{Terms of second order in $u_{\rm_S}$.}
\begin{tabular}{lcccc}
\hline
Var.  & Eq. & $\mathcal{O}(u_{\rm s}^2)$ &  Value\\
\hline
a & \ref{eq:modulation_amplitude}   & $(\tan^2\alpha + \frac{1}{6})u_{\rm s}^2$  & 0.026 \\
$\mathcal{F}_x$    & \ref{eq:sfthrustx}    &  $-(\tan^2\alpha - \frac{1}{6})u_{\rm s}^2$   & -0.019 \\
$\mathcal{F}_z$   & \ref{eq:sfthrustz}  &  $(\frac{3}{4}\tan^2\alpha - \frac{1}{3})u_{\rm s}^2$   & 0.009 \\
$\mathcal{F}_{\parallel}$   & \ref{eq:sfthrustpar}  &  $\frac{1}{2}(\tan^2\alpha - 1)u_{\rm s}^2$   & 0.000 \\
$\mathcal{F}_{\perp}$   & \ref{eq:sfthrustperp}   & $\tan^2\alpha u_{\rm s}^2$   & 0.023\\
$\tan\psi$ & \ref{eq:tanpsi} & $-\frac{(3\tan^2\alpha + 2)u_{\rm s}^2}{6(2-\sin^2\alpha)}$  & -0.013 \\
\hline
\end{tabular}
\begin{tablenotes}
      \small
      \item Values of the second order terms are evaluated at $\alpha = 45^{\circ}$ and $u_{\rm s} = 0.15$.
    \end{tablenotes}
\label{tab:2ndorder}
 \end{threeparttable}
\end{table}

\begin{figure}[htb]
	\centering\includegraphics[width=2.7in]{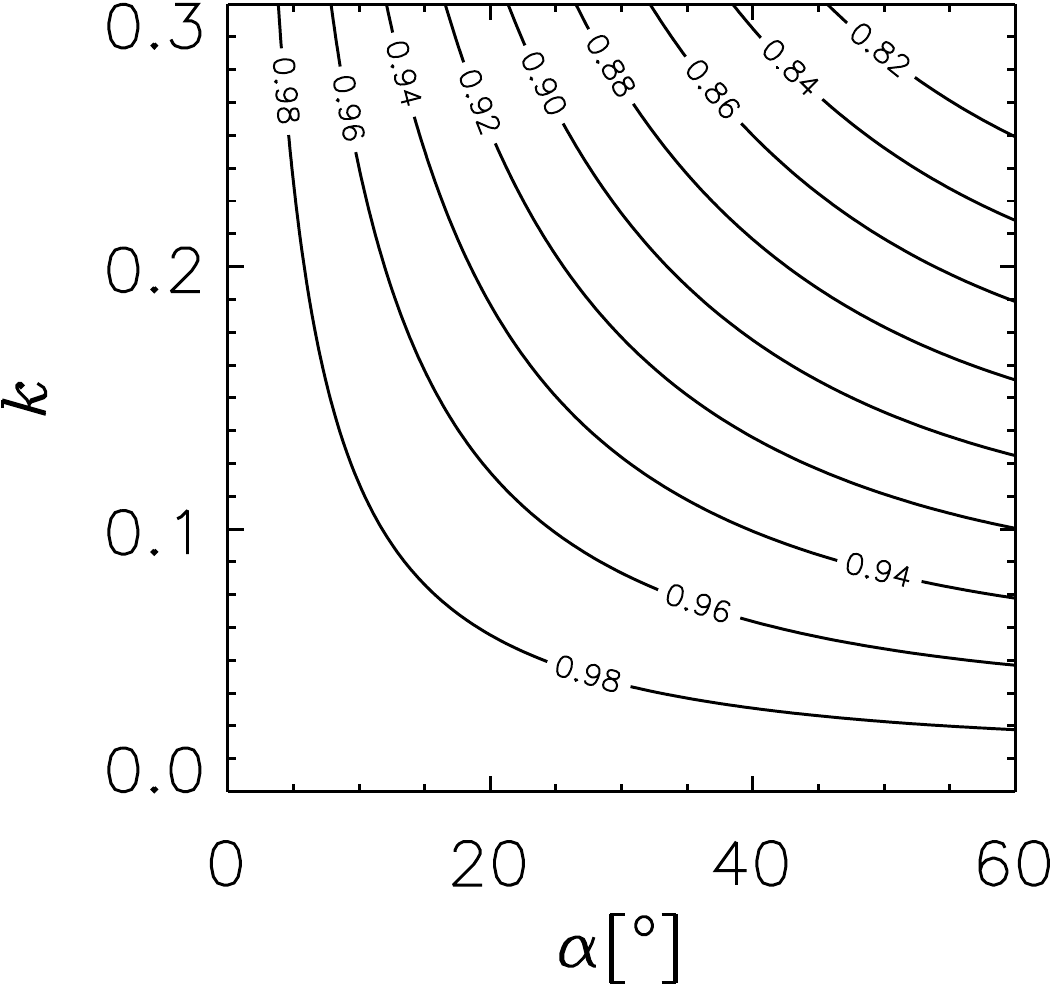}
	\caption{Efficiency of the tether voltage modulation.}
	\label{fig:efficiency}
\end{figure}

As a comparison, for a rigid tether model without auxiliary tethers,
the modulation amplitude equals $3\tan\Lambda\tan\alpha$
\cite{controla}, where $\Lambda$ is the rigid tether coning angle. The
percentage difference between these two models is shown in
Fig. \ref{fig:delta_eff}. For this model, the angular velocity of the
tether varies as the tethers are not mechanically coupled, and the
tether angular velocity varies over the rotation phase enhancing the
amplitude of the voltage modulation. Also a model with rigid tethers
and auxiliary tethers can be considered (the sail resembles the Asian
conical hat). The analysis of such a model is similar to the one
carried out in this paper, and the modulation amplitude for such a
model equals $2\tan\Lambda\tan\alpha$. It can be seen that both the
mechanical coupling and the realistic tether shape increase the sail
efficiency as shown by Eq. (\ref{eq:modulation_amplitude}).

\begin{figure}[htb]
	\centering\includegraphics[width=2.7in]{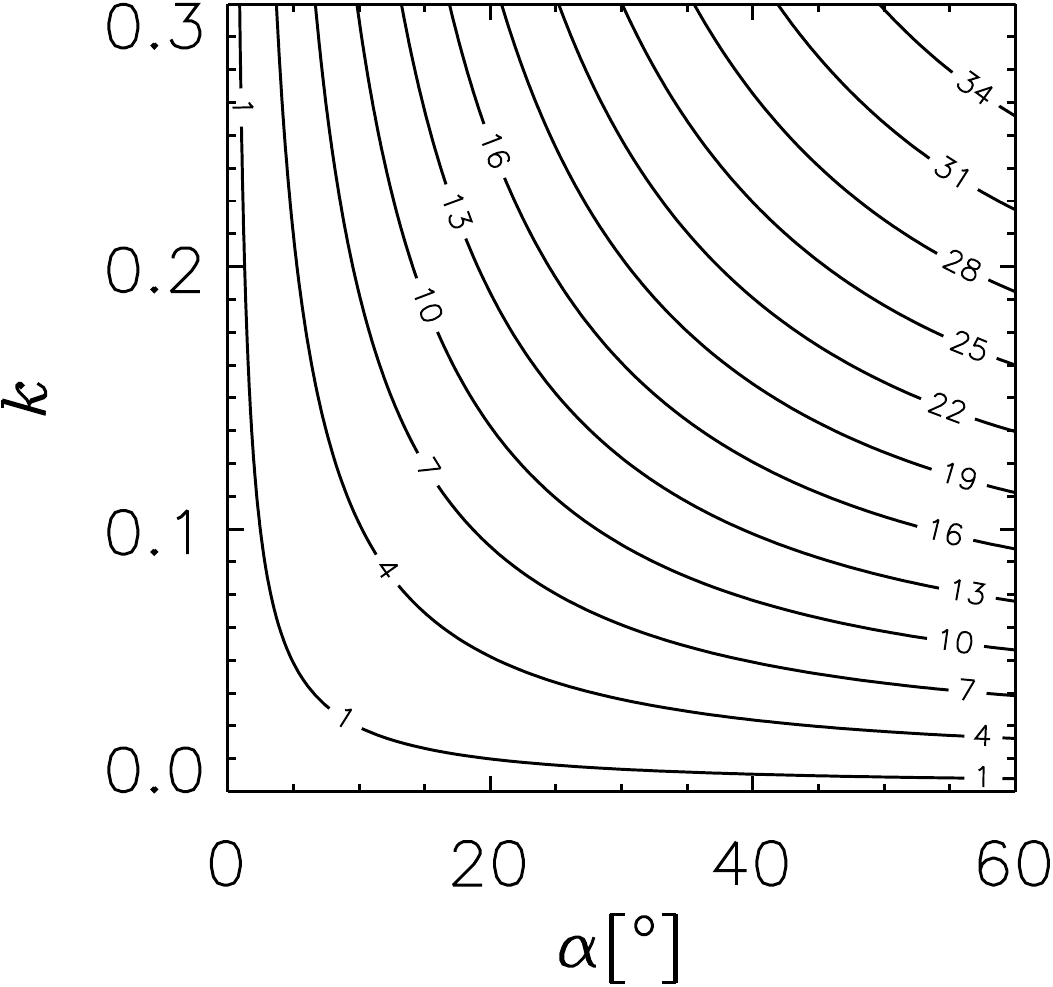}
	\caption{Percentage difference in sail efficiency between the realistic sail model and a single mechanically uncoupled tether model \cite{controla}.}
	\label{fig:delta_eff}
\end{figure}

\subsection{Thrust vectoring}
\label{subsec:thrustvectoring}

Using the voltage modulation (\ref{eq:modulation}) in
Eq. (\ref{eq:totalforce}), the total thrust can be integrated over the
tethers in the case of the torque-free sail flight orientation determined
by the sail angle $\alpha$,
\begin{eqnarray}
\mathcal{F}_x &=& \mp\frac{1}{2} N \xi L v\sin\alpha \left(1 + u_{\rm s}\tan\alpha + \mathcal{O}(u_{\rm s}^2)\right) \label{eq:sfthrustx}\\
\mathcal{F}_z  &=& - N \xi L v\cos\alpha \left(1 + u_{\rm s}\tan\alpha + \mathcal{O}(u_{\rm s}^2)\right).
\label{eq:sfthrustz}
\end{eqnarray}
The thrust components can then be rotated by the sail angle $\alpha$
to give the transverse and radial thrust components as
\begin{eqnarray}
\mathcal{F}_{\parallel} &=& \pm\frac{1}{4} N \xi L v\sin 2\alpha\left(1 + u_{\rm s}\tan\alpha + \mathcal{O}(u_{\rm s}^2)\right) \label{eq:sfthrustpar}\\
\mathcal{F}_{\perp} &=& -N \xi L v\left(1 - \frac{1}{2}\sin^2\alpha\right)\left(1 + u_{\rm s}\tan\alpha + \mathcal{O}(u_{\rm s}^2)\right).
\label{eq:sfthrustperp}
\end{eqnarray}
Fig. \ref{fig:transverse} shows the dimensionless transverse thrust
component of the sail thrust. Naturally, the transverse thrust is
enhanced as the sail angle increases reaching the maximum of about one
fourth of the total electric sail force at $\alpha = 45^{\circ}$. As a
comparison, the decay of the transverse thrust in $k$ is somewhat
slower than that of the single tether model. This is clarified in
Fig. \ref{fig:delta_trans} that shows the percentage difference in
transverse thrust magnitudes between these two models.
\begin{figure}[htb]
	\centering\includegraphics[width=2.7in]{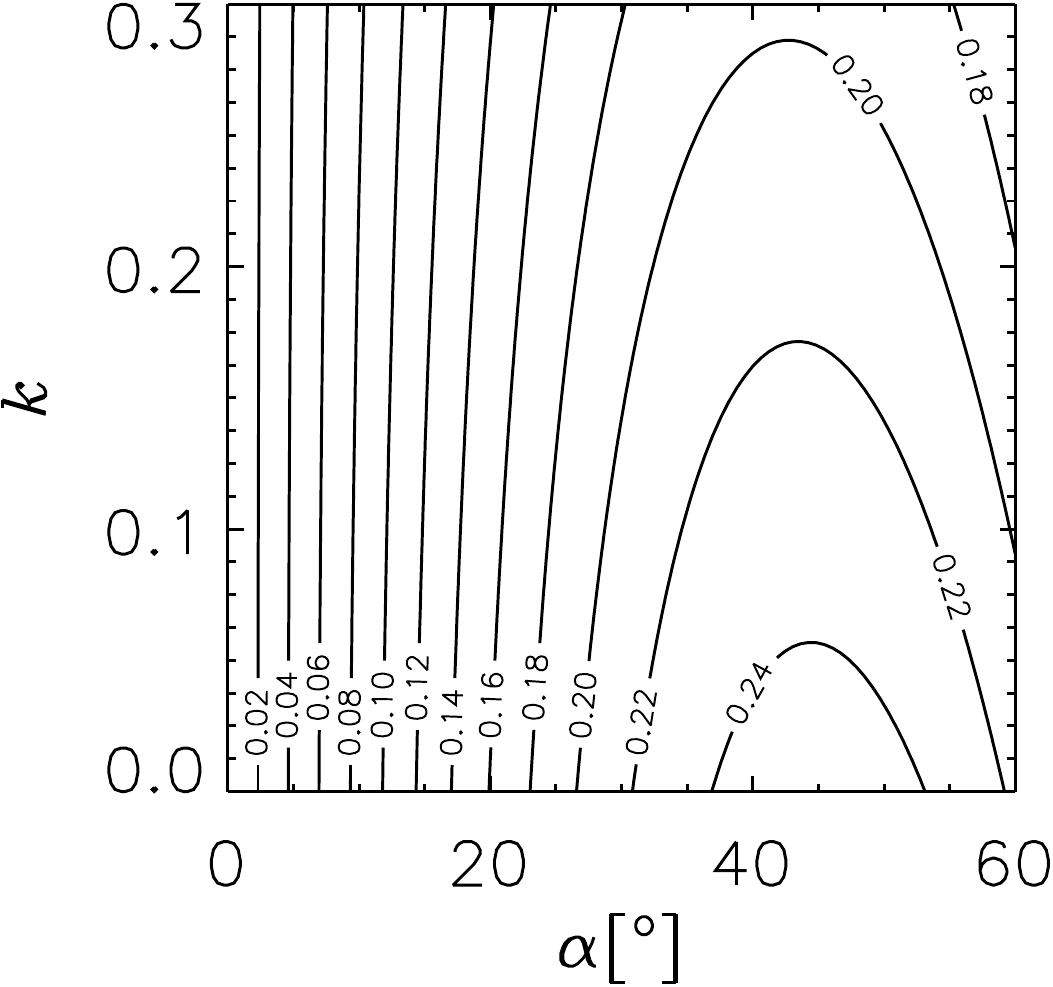}
	\caption{Transverse thrust component normalized to the maximum
          available electric sail force.}
	\label{fig:transverse}
\end{figure}
\begin{figure}[htb]
	\centering\includegraphics[width=2.7in]{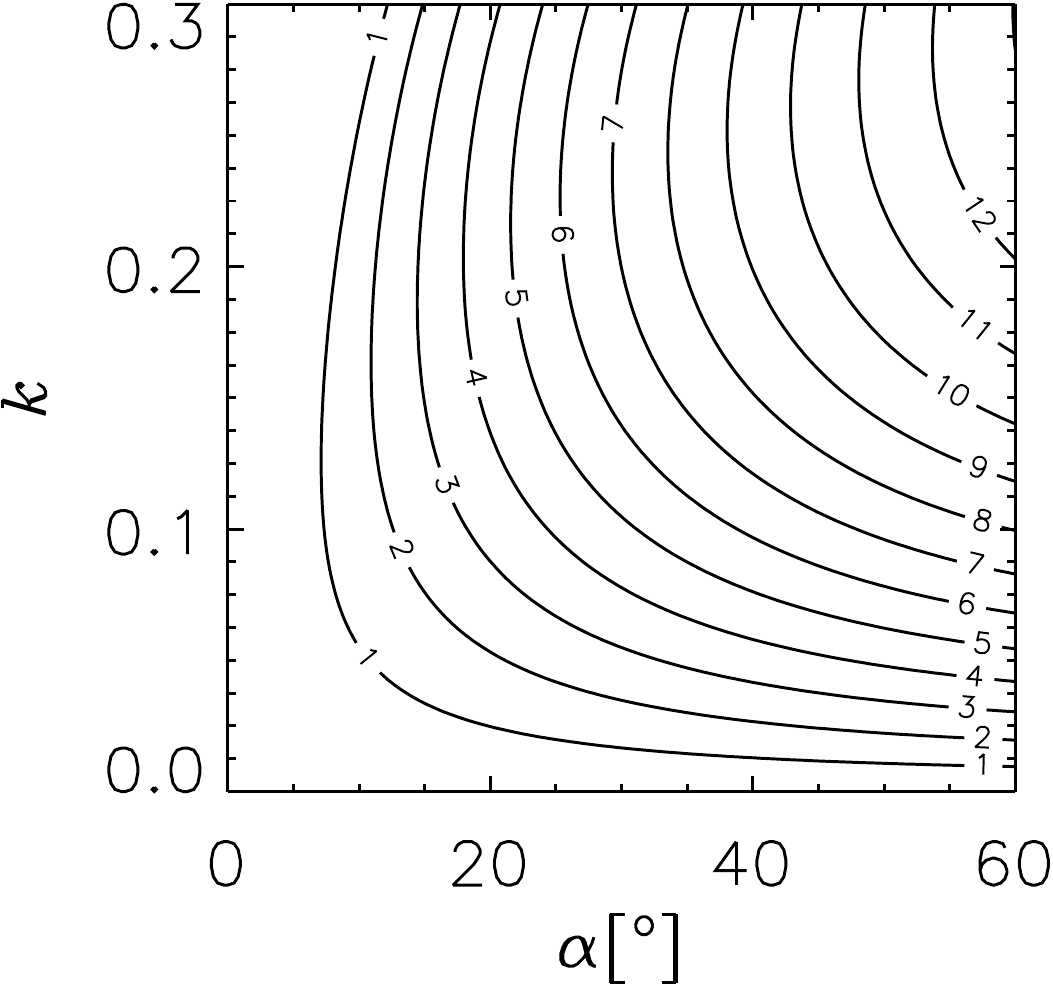}
	\caption{Percentage difference in transverse thrust between the realistic sail model and single mechanically uncoupled tether model.}
	\label{fig:delta_trans}
\end{figure}

Finally, the tangent of the thrusting angle can then be written as
\begin{equation}
\tan\psi = \mp\frac{\sin 2\alpha}{2(2-\sin^2\alpha)}\left(1 + \mathcal{O}(u_{\rm s}^2)\right).
\label{eq:tanpsi}
\end{equation}
It can be seen that the thrusting angle (Fig. \ref{fig:thrustinga})
has only a weak dependence on the sail root coning tangent $u_{\rm
  s}$. Thus the thrusting angle can be computed by assuming that the
sail is fully planar ($\tan\psi = \mp\sin 2\alpha/(4-2\sin^2\alpha)$).
\begin{figure}[htb]
	\centering\includegraphics[width=2.7in]{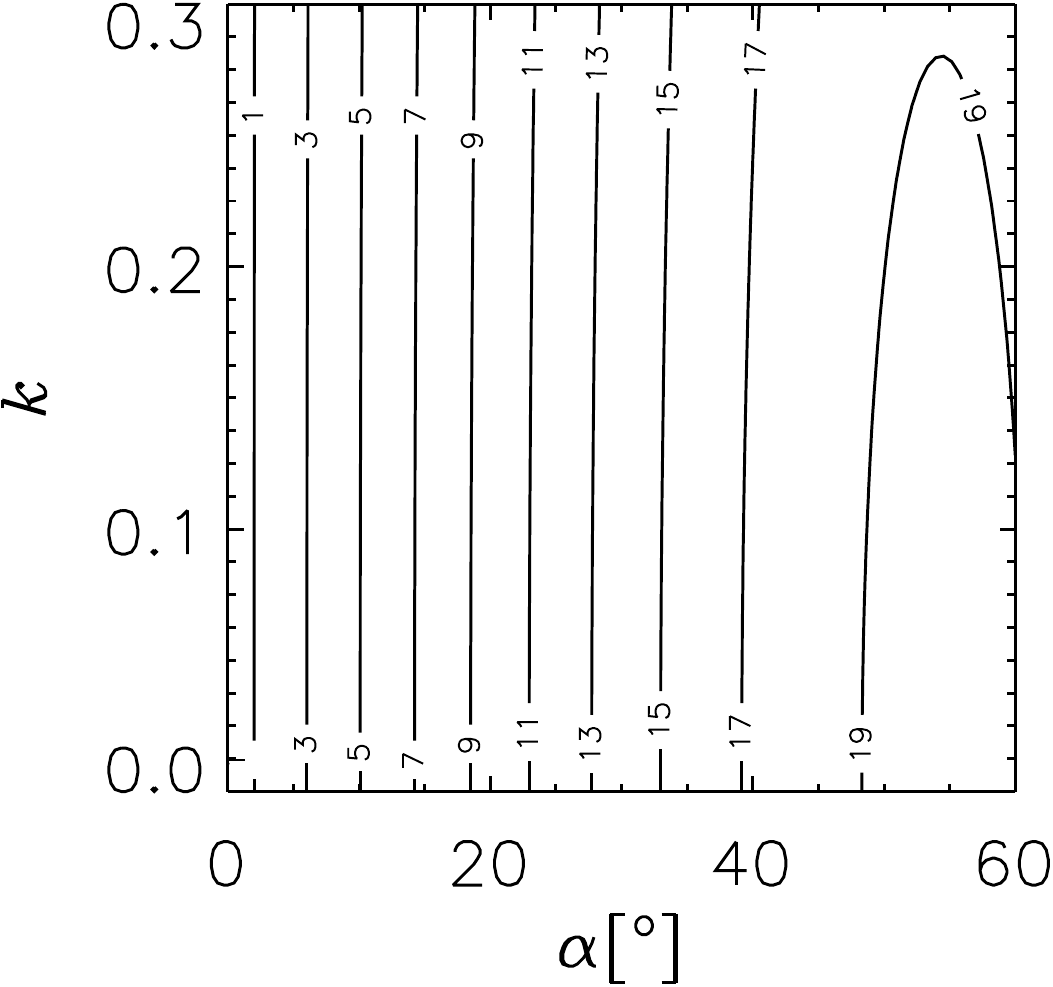}
	\caption{\cra{Thrusting angle given in degrees.}}
	\label{fig:thrustinga}
\end{figure}


\section{Discussion and conclusions}
\label{sec:summary}

In this paper, we assumed that the solar wind is nominally flowing
radially from the sun. This served the purposes of this paper which
was to estimate the effects of the actual sail shape to the efficiency
of the sail control and thrust vectoring. When solar wind temporal
variations are considered, the $y$ component of the solar wind must be added
in the sail torque components in order to write a complete rigid body
simulation for the electric solar wind sail. Furthermore, the Euler
equations require also the moments of inertia in addition to the
torques given in this paper. However, both the general thrust
components in the sail body frame and moments of inertia can be
attained with a reasonable effort by following the analysis of this
paper, especially, when using a computer algebra. Such a complete
Euler description of the electric solar wind sail can then be used,
for example to address the effects of the solar wind variation to the
sail navigation, and spin rate control and evolution in sail
orientation maneuvers.

In this paper, we derived the equation of tether shape,
solved it by a simple numerical iteration, and presented an analytical
approximation for the single tether shape. Our approximation is
parametrized by the tether root coning angle and the tether
length. The latter is a free parameter whereas the former depends both
on the ratio of the electric sail force to the centrifugal force and
the sail angle with respect to the sun direction. This ratio then
depends on the tether voltage, solar wind density and speed, sail spin
rate, and total mass of the tether and remote unit combined. The sail
coning angle at the spacecraft is essentially the tether root coning
angles averaged over the tether locations in the sail rig. The
resulting sail shape is such that the coning decreases and the sail
surface tangential to the tethers approaches the sail spin plane
towards the perimeter of the sail.

Having obtained the model for the sail, we derived expressions for the
angular thrust and torque densities. Introducing a tether voltage
modulation that results in torque-free sail dynamics, we solved the
amplitude of the modulation. This amplitude has to be reserved for the
sail control and correspondingly the voltage available for thrusting
is less than the maximum designed voltage increasing the sail
efficiency. We showed that this amplitude is 3 times smaller for the
sail model introduced here than for that derived using a single tether
model \cite{controla}. Finally, the total thrust to the sail was
obtained for the torque-free sail motion. The transverse thrust is
somewhat larger (up to about 10\%) than that of the single rigid
tether model. The reason is that a portion of the sail near the
perimeter of the sail is coplanar with the sail spin plane. The
thrusting angle was shown to be essentially equal to the fully planar
sail being about 20$^{\circ}$ at sail angles higher than 45$^{\circ}$.




\section*{Acknowledgments}
This work was supported by the Academy of Finland grant 250591 and by
the European Space Agency.

\bibliographystyle{model6-num-names}
\bibliography{<your-bib-database>}



\end{document}